\documentclass{article}
\usepackage{graphicx}
\usepackage{indentfirst,csquotes}

\usepackage{amssymb,amsthm,amsmath}
\usepackage{xcolor,paralist,hyperref,fancyhdr,etoolbox}

\usepackage{natbib}
\usepackage{float}
\usepackage{array}
\usepackage{tabularx}
\usepackage{booktabs}
\usepackage{multirow}
\usepackage{changepage}
\usepackage{fullwidth}
\usepackage{fancyhdr}

\fancypagestyle{FirstPage}{
\lfoot{Parts of this manuscript have already been published in: Benedict Witzenberger and J{\"u}rgen Pfeffer. 2023. Gender Dynamics of German Journalists on Twitter. In Proceedings of the 2022 IEEE/ACM International Conference on Advances in Social Networks Analysis and Mining (ASONAM '22). pp. 226–230. \textcopyright IEEE Press 2023 https://doi.org/10.1109/ASONAM55673.2022.10068698} 
}

\newlength{\extralength}
\providecommand{\keywords}[1]
{
    \small
    \textbf{\textit{Keywords---}} #1
}

\newcolumntype{C}{>{\centering\arraybackslash}X}

\hypersetup{ colorlinks=false } 

\graphicspath{{img/}}

\usepackage{comment} 

\begin{document}
\title{More inclusive and on wider sources: A Comparative Analysis of Data and Political Journalists on Twitter in Germany}

\author{Benedict Witzenberger and J{\"u}rgen Pfeffer}
\date{School of Social Sciences and Technology, Technical University of Munich, Germany\\[2ex]
\today}

\maketitle

\abstract{Women are underrepresented in many areas of journalistic newsrooms. In this paper, we examine if this established effect continues in the new forms of journalistic communication, Social Media Networks. We used mentions, retweets, and hashtags as journalistic amplification and legitimation measures. Furthermore, we compared two groups of journalists in different stages of development: political and data journalists in Germany in 2021. Our results show that journalists regarded as women tend to favor their other women in mentions and retweets on Twitter, compared to men. While both professions are dominated by many men and a high share of men-authored tweets, women are mentioning and retweeting other women to a more extensive degree than their male colleagues. Women data journalists also leveraged different sources than men. In addition, we have found data journalists to be more inclusive towards non-member sources in their network compared to political journalists.}


\section{Introduction}
\thispagestyle{FirstPage}

Social Media Networks (SMNs) such as Twitter have had a significant impact on journalism. Researchers have focused on how Twitter has challenged key values of journalism, such as objectivity, gatekeeping, and transparency \citep{Lasorsa2012a,Lawrence2014,Hermida2010}. Twitter and other microblogging platforms have also changed the news cycle by creating a hybrid system of new actors and news-sourcing habits \citep{Chadwick2011}. Journalists commonly use Twitter as a source of information \citep{Paulussen2014}. Some have noted changes in the way private and professional personae are presented on Twitter, which may collide with corporate brands \citep{Ottovordemgentschenfelde2017,Hanusch2018}.

A rigid selection of information shapes the world of SMNs. This is not a new development. \citet{Lippmann1946} described the bias between reality and perception — or mental image — around 100 years ago as a pseudo-environment. This explains the selective way of processing information shaped by social constructs surrounding the individual that has been researched since then \citep{Lazarsfeld1944}.

A sender-based selection form was described by \citet{Lewin1947}, showing that disseminators tend to spread information that aligns with their values. The foundation of the gatekeeping theory \citep{White1950} has shaped journalism over decades but has become a more general phenomenon since the global spread of information is no longer restricted to journalists but open to everyone on a social media platform. This led to an increase of data that might help to shed light on processes that have so far taken place behind closed doors. In this article, we try to enhance our understanding of journalistic discourses on Social Media, focusing mostly on gender and journalistic areas as differentiators.

The history of women's journalism is much older than Social Media. Female journalists were first hired in the second half of the 19th century out of financial interests: they were needed to help create so-called ``women's pages'' \citep{Steiner2008,Hunter2019,Chambers2004,Kay2012} with topics like fashion, art, or societal gossip. These "women's pages" targeted a female audience that the newspapers wanted to attract because of the increasing revenues from advertising in the newspapers \citep{Lang1999}. Currently, women journalists are underrepresented in many newsrooms. They are therefore less visible in the media \citep{Smith1981,Hannis2007,Kian2009,North2016}, which might lead to distorted ``news-is-for-men'' perceptions in the audience \citep{Sui2022}. Other channels of public appearance could provide new platforms for women journalists to promote their work or build reputations in their beats. Twitter, as a platform with few barriers to entry, naturally would be expected to serve as an enhancement to building a platform. However, previous work has shown that this is not necessarily the case \citep{Lasorsa2012,Usher2018}; political journalists especially have been shown to form male-dominated, elitist networks \citep{Matusitz2012,Lawrence2014}.

We build on an emerging body of literature that uses Twitter data to analyze networks of journalists to find out if there are sex-related differences between journalists on Twitter in general and groups of journalists in particular. We focus on two groups: political and data journalists in Germany. Journalists, as a profession, play an important role in disseminating information and shaping public opinion. However, this is most visible for political journalists who often cover issues of profound societal and political significance. Their presence on social media platforms like Twitter can have policy implications and influence public discourse on critical topics. Investigating the behaviors of political journalists on Twitter contributes to a broader understanding of the interplay between journalism, politics, and society, which has become increasingly important in the digital age. 

Data journalism, in particular, has witnessed significant growth and innovation in recent years. Data visualization, interactive storytelling, and data-driven investigations have become increasingly prevalent in journalism. By focusing on data journalists, we aim to capture emerging trends in journalism practice and explore how these innovations manifest on Twitter.

We analyzed 478,263 tweets from political and data journalists in Germany in 2021 to compare communication styles within these communities and between sexes. Men dominate the number of tweets, whereas women tend to favor other women in mentions in general and retweets of political journalists. We also find men to be self-retweeting themselves a lot more than women, and different sourcing behavior in both groups of journalists. 

The contributions of this article are:

\begin{itemize}
    \item We provided a first comparative quantitative analysis of communicative differences between and within German political and data journalists, offering a non-US-centric perspective.
    \item We found manifestations of existing sex-related norms in the Twitter behavior of both journalist groups, confirming prior studies on US-political journalists.
    \item Our analysis of sources and hashtags revealed a broader spectrum of sources for data journalists and a different sharing behavior of men and women.
\end{itemize}

We will start our argument by laying out related literature on gender issues in journalism, then give an overview of political and data journalists in Germany, followed by methods and a result section.

\section{Literature Review}

Prior work has studied women in journalism and on SMNs, as well as the behavior of political journalists on Twitter.

\subsection{Journalism and Gender}

Journalism and gender have been studied from various perspectives during the last two decades. Most work focuses on a comparative perspective and a possible influence on journalistic style between men and women \citep{Craft2004,Hannis2007,Kian2009,North2016}.

While parity for lower-level editorial jobs has been nearly achieved, there is still a discrepancy in higher-level jobs in the newsroom \citep{Ziamou2000,Byerly2011,Chambers2004,Andi2020}. This ``glass ceiling'' impacts the paper policy as higher-ranking positions proclaim the editorial ethos. This leads to a limited perspective on issues \citep{Smith1981,Fleras2003} and differences in attributing beats \citep{Craft2004}. 

Several studies on the size of the gender gap in Germany, all with slightly different methods, and none with current figures exist. A tally by the initiative ``ProQuote,'' which lobbies for a women quota of 30 percent, found shares of women in power (including editors-in-chief down to deputy leaders of sections) between 16.1 and 50.8 percent for national newspapers in June 2019 \citep{Garmissen2019}, totaling to 25.1 percent women versus 74.9 percent men. Across all positions, the ``Worlds of Journalism Study'' in 2016 found a proportion of 40.1 percent women \citep{Hanitzsch2016,Steindl2017}, while a 2013 study by the European Institute for Gender Equality, which only included a few media corporations, estimated around 44 percent \citep{GenderEqualityEIGE2013}.

\subsection{Twitter for Journalists}

Since its creation in 2006, Twitter and its implications on journalism have been studied in multiple dimensions, as previously mentioned. It has primarily been described as a platform for breaking news \citep{Kwak2010}, with its users mainly talking about headlines and current affairs \citep{Kwak2010,Asur2011}. Twitter is a medium for professional communicators --- like politicians and celebrities. This aspect of Twitter seems to make it more appealing to journalists in comparison to other professions \citep{Nordheim2018}.

This is especially true of political journalists: The platform is, if not a central source of news in Washington D.C. \citep{Hamby2013,Kreiss2016}, a central source of news in several Westminster democracies \citep{Hanusch2018} and Germany \citep{Nuernbergk2016,Nuernbergk2020,Degen2021}.

Over time, the use of Twitter has increasingly normalized for journalists \citep{Lasorsa2012a}. However, this process takes time and requires corporate policies that prevent the fast, widespread adoption of newer features \citep{Molyneux2019}. Others have described this process less as normalization and more as a negotiation between traditional gatekeeping roles and editorial decision-making and the new influences injected by users on SMNs \citep{Tandoc2016}.

While most studies concentrate on the implications of the social network on journalistic work, several authors have investigated the effects and foundations of journalistic networks on Twitter on the journalists themselves.

Verweij analyzed Twitter links and communication networks between 150 Dutch journalists and politicians and compared in- and out-degrees. As a result, two groups of top actors emerged: news sources with high in-degrees, which are highly connected and form a ``political Twitter elite'' \citep[p.~685]{Verweij2012} and news gatherers with high out-degrees who search for information. Both groups act as bridges and links in the network.

Multiple studies have shown political journalists to form elitist circles on Social Media. Research on the tweeting behavior of the reporters covering the 2012 Republican and Democratic conventions showed that journalists tend to express more opinions in their writing on Twitter than in journalistic media. A sample on a list of 430 reporters and commentators was drawn and manually coded. Reporters consistently kept a closed gate-keeping level by linking and retweeting mainly themselves and their fellow reporters and rarely reacting to their followers \citep{Lawrence2014}. 

Further research on a similar dataset of the 2012 presidential race showed that reporters kept their tweets focused on the main topic and rarely questioned their peers' views but used Twitter as a ``space for collective interpretation of political events'' \citep{Mourao2015}. This view describes the journalists as forming a virtual ``bubble,'' or an ``interpretive community'' \citep{Zelizer1993}.


Evidence from the 2016 presidential race in the U.S. suggests similar results, although the study was limited to retweets, quote tweets, and replies \citep{Molyneux2019}. This observation seems to hold even after Twitter's user base was stabilized. Several other scholars have shown that journalists discuss issues mostly with other journalists or politicians \citep{Mourao2016,Maares2021}

Others have used the concept of homophily to estimate the closeness of journalistic networks. Australian journalists have been found to share a significant degree of homophily in characteristics like gender, organization, and geographic proximity. The largest amount of homophily is attributed to their beat. This leads to a tightly-knit, homogenous, elitist community, largely interacting with itself \citep{Hanusch2019}. This strongly supports other findings that showed that political journalists prefer to engage with other political journalists, whether in opinion-building on political debates or joking about the debates.

\subsection{Twitter and Gender dynamics}

The behavior of journalists in SMNs has been part of several studies. On the one hand, women journalists on Twitter reveal more about their personal lives and link more to external websites, indicating more transparency than their male peers \citep{Lasorsa2012}. An analysis of political reporters in Washington, D.C., showed that male journalists amplify and engage their male peers nearly exclusively. Women also engage with each other but retweet men more in absolute terms than retweet women \citep{Usher2018}. On the other hand, women journalists frequently encounter sexual harassment in online environments \citep{Stahel2020}, especially when covering topics that are somewhat regarded as men's territory \citep{Sarikakis2021}. This has been shown to limit their ability to communicate with their audience \citep{Chen2020}, lead to avoidance \citep{Adams2018,Stahel2020}, and is described to be aimed at disciplining journalists \citep{Waisbord2020}.

\subsection{Twitter use of political journalists in Germany}

For reporters in the German parliament, the Bundestag, Twitter is the most used social media network for journalists covering federal politics in Berlin used to observe sources and topics and gather information \citep{Nuernbergk2016,Nuernbergk2020}. 

Research has suggested that the interpretive standpoints chosen in their reporting can already be concluded by looking at the tweets of political journalists \citep{Degen2021}. Furthermore, Twitter interactions between politicians and journalists can lead to different assessments of Twitter, compared to journalists with no interactions --- which indicates that the network also plays a role in relationship management \citep{Nuernbergk2020}.

Research from 2014 has shown that the correspondents include politicians in their communicative circles but stick together when debating, not reacting to other users who try to contribute to the discussion \citep{Nuernbergk2016}. This is consistent with other authors, as previously above.

\subsection{Data journalists in Germany}

Data journalism is a newer playfield of journalism. While its roots are mostly dated back to the 1970s idea of ``precision journalism'' \citep{Meyer1973,Meyer2002,Coddington2015,Bravo2020}, some even go as far as defining its main provenance in the works of or tables in the Guardian in 1821 or visualizations by Florence Nightingale and Jon Snow in the 1850s \citep{Rogers2010}; however, it is mainly regarded to have been started around 2009 \citep{Bravo2020}. Its main focus is the combination of data analytical approaches to find and extract information in data and tools to visualize the results and tell stories with it, which enhance traditional reporting \citep{AndertonYang2012,Coddington2015,Berret2016,Antonopoulos2020}. 

Data journalism in Germany has been enumerated twice: In the spring of 2013, Weinacht and Spiller \citep{Weinacht2014} identified 35 individuals working as data journalists in Germany and were able to interview them, and in 2020 Beiler et al. \citep{Beiler2020} estimated that data journalism is well established in three-fourths of media outlets.

While there is no published data on the gender share of data journalists in Germany, a look at the 2013 study of Weinacht and Spiller, which intended to cover all data journalists in the country at that time, reveals that 3 out of 35 interviewees have a women first name \citep{Weinacht2014}. Likewise, a study on data journalists in Sweden in 2014 found that 46 percent of the respondents were women, 53 percent were men, and two percent declined to answer \citep{Appelgren2014}.

Compared to other areas, data journalism is regarded as a new field not guarded by ``old boys'' networks, thus being more open to all genders \citep{Vuyst2018}. This allows data journalism access to journalistic areas formerly more exclusive, like investigative reporting. On the downside, there is a lack of women in technical positions, which spills over into a lack of women in data journalism because they lack the skills to apply. This is seen as a lack of women in computer sciences \citep{Vuyst2018}. In a self-assessment study, men data journalists rated themselves as more experienced than women \citep{Appelgren2014}, although it is unclear if this is due to men's overconfidence or the understatement of women.

\subsection{Hypotheses}

To structure this research, we present three questions, split into five hypotheses we aim to answer.

\subsubsection{The ``boys on the bus'' are now on Twitter}

The first hypothesis is centered around the idea of an elitist community of political journalists found multiple times in the past \citep{Lippmann1946} --- Twitter could, by default, have an opening effect on those groups. Political journalists have been shown to form elitist circles on Social Media \citep{Lawrence2014,Mourao2015,Nuernbergk2016,Molyneux2019}. We want to compare them to data journalism as a newer form of journalism. Because the latter is derived from a more technical, computer science-driven background --- so-called ``programmer-journalists'' \citep{Parasie2012} --- they may have different approaches to communication. Data journalism is often regarded as more transparent in its underlying data and methods \citep{Diakopoulos2016}, which might be conveyed differently in Social Media discourses. Our first research question is as follows:

\textbf{RQ1}: Are data journalists engaging differently with non-peers on Twitter than political journalists?

The primary hypothesis is:

\textbf{H1:} Data journalists have a more open discourse than political journalists.

\subsubsection{Journalistic gender dynamics on Twitter}

Another set of hypotheses is aligned with the question of gender dynamics in the Twitter behavior of journalists.

Twitter plays a vital role in publicly providing journalistic legitimation \citep{Carlson2017} or dominance in a specific field \citep{Barnard2014}. This has also been argued above when excluding outsiders from discourses but is also true within the field when establishing a hierarchy \citep{Mourao2015}.

As \citet{Usher2018} have shown, there has been men's dominance in Washington D.C.'s political journalism's use of Twitter. Not only do male journalists amplify their gender, but women also tend to mention and retweet male correspondents more than their peers in absolute terms. Relatively, women retweet other women much more than expected by the raw share of genders. Selective behavior as an inherent behavior in SMNs has already been described earlier by showing that men retweet primarily men and women retweet mostly women \citep{Xiao2012}. In the case of journalism, the quantity of the observed gender gap is striking, being described as a ``gendered echo chamber'' \citep[p.~338]{Usher2018}.

These results were retrieved by calculating so-called power users for typical Twitter activities, attributed to specific categories: replying or following was a measurement for engagement, mentioning for legitimation, and retweeting and quoting for amplifying. Our analysis uses mentions and retweets as indicators for legitimizing or amplifying behavior.

\textbf{RQ2}: Are there differences in gender bias in mentions and retweets of German politics and data journalists on Twitter?

We created two hypotheses for our Tweet analysis:

\textbf{H2}: Women journalists are mentioned less than men.

\textbf{H3}: Women journalists are retweeted less than men.

\subsubsection{Differences in sources}

A third perspective is based on the content of the tweets that are shared by both sexes in the studied journalistic disciplines. Earlier research has suggested that women journalists tend to be assigned to types of stories regarded as being `soft', like arts, education, or health \citep{North2016}, and use different sources in their reporting \citep{Armstrong2004}. As this research already focuses on a narrow subset of journalism, we want to understand if these observations hold true on Twitter, making it easier for journalists to elevate sources and focus on topics important to them without having to clear editorial processes.

As data journalism is derived from a very broad set of backgrounds, we would expect them to leverage a more diverse set of sources than political journalists.

We therefore ask:

\textbf{RQ3}: Can we identify differences in retweeted sources or hashtags between sexes and German political and data journalists on Twitter?

To answer this question, we raise two hypotheses:

\textbf{H4}: Women journalists amplify different sources and hashtags than men.

\textbf{H5}: Data journalists have a more diverse set of topics than political journalists.

\section{Materials and Methods}

To provide an accurate and detailed snapshot of \textbf{German political journalists} on Twitter, we based the selection on the circulation and size of German newspapers. We tried identifying journalists who are clearly deployed to political sections or mainly working on political topics. This approach limits the proportion of regional newspapers, which use news agencies more extensively in their political reporting than larger newspapers and have no apparent political reporters. Many larger newspapers offer an imprint with an overview of their authors and their vitae, which often contains Twitter accounts. Smaller newspapers sometimes lack that information, which must be retrieved from the articles. From these 730 accounts, all Tweets between January 1 and December 31, 2021, were retrieved on January 7, 2022, using Twitter API v2 \citep{pfeffer2023}---in total 430,451 tweets.

Journalists working for T.V. or radio stations- largely public corporations in Germany- have been omitted. The importance of newspapers has been assessed by two publications: the quarterly circulation data provided by the so-called "Informationsgemeinschaft zur Feststellung der Verbreitung von Werbetr{\"a}gern e.V." (abbreviated IVW) --- which corresponds to the "Audit Bureau of Circulation" --- and the recurring "Media-Analyse," a research survey that tries to evaluate the media consumption habits of the German population. We used data from 2019 for regional and 2020 for national newspapers to determine the publications needed to look further for journalists who use Twitter.

This approach differs from Nuernbergk's \citep{Nuernbergk2016}, which used a predefined set of political journalists who are members of the official German Federal Press Conference. As a result, we expected our sample to include more journalists located in areas other than the German capital of Berlin.

Accounts that were obviously only private --- meaning they showed no connection to the newspaper or regularly mentioned its stories --- were discarded. This list was compiled in July 2020 and updated on January 7, 2022. 
 
To identify \textbf{data journalists}, we used an advocacy group as a starting point. A majority of German data journalists have decided to congregate as a so-called ``Fachgruppe'' (\textit{professional group}) of the non-governmental reporters' representation "Netzwerk Recherche" in the fall of 2020, "Netzwerk Recherche" sees itself ``as general representatives of the interests of the entire field of data journalism and all its manifestations'' \citep{NetzwerkRecherche2020}. 
To simplify communications, a group on the messaging platform Slack was created, open to anyone who regards herself or himself as a data journalist. The restriction on this platform introduces a form of self-selection, which may lead to bias in this research. However, we assume the majority of data journalists to be members of this group, as there are no fees or further barriers, and participation in the group offers incentives, like discussions on current topics in the field, information on upcoming conferences or meet-ups, or a job market.
We acknowledge potential privacy concerns introduced by using a somewhat non-public data source. We did, however, not analyze data on an individual level.
We identified 167 members at the time of our data collection, similar to what has been collected in previous studies \citep{Beiler2020,Haim2022}. Twitter usernames and affiliations were manually added whenever mentioned in the profile's description text. 148 data journalists could be connected with a Twitter account, and 47,812 tweets were downloaded on March 11, 2022, for 2021. See Table~\ref{tab:gender_shares} for comparing extracted numbers.

\begin{table}[H] 
\caption{Comparing shares of sexes in our sample by group of journalists with a total from the ``Worlds of Journalism Study'' \citep{Steindl2017}.\label{tab:gender_shares}}
\begin{tabularx}{\textwidth}{CCC}
\toprule
          & \textbf{Women}  & \textbf{Men}    \\
\midrule
Total     & 40.1\% & 59.9\% \\
Political & 28.6\% & 71.4\% \\
Data      & 32.2\% & 67.8\% \\
\bottomrule
\end{tabularx}
\end{table}

\subsection{Adding a gender attribution.}

We assigned a binary gender category to all users in our lists by manually coding the author's first name into traditional male or female first names. This approach may result in misspecifications if someone identifies as a different gender, as expected by the name. It has to be noted that this reliance on a binary gender framework may also not adequately capture the complexities of gender identity. It may exclude non-binary and transgender journalists, whose interactions on social media could offer valuable insights into the broader conversation about gender dynamics in journalism.

However, as this work tries to identify a potential divergence between users who appear as women and men for outsiders and aims to be comparable to prior studies, we consider this issue approach sufficient. In unclear cases, we tried to deduce the gender using profile pictures.

No names have been found which were not explicit enough to be assigned to a gender. In our data, 28.6 percent of political and 32.2 percent of data journalist users were regarded as women, while 71.4 percent of political and 67.8 percent of data journalists were identified as men. Both groups have fewer shares of women than the ``Worlds of Journalism Study'' found in Germany in 2016, with 40.1\%.

\subsection{Clustering sources and hashtags}

Incorporating the clustering of retweet sources and hashtags into our study constitutes an approach that enriches the depth of our analysis by providing a more comprehensive understanding of the content of tweets within the context of journalistic communication. 

We first extracted all retweeted usernames throughout our dataset, totaling 20,937 accounts for political journalists and 6,519 for data journalists. After removing self-retweets, we extracted the 30 most retweeted accounts for both groups and both genders of journalists and labeled them into categories (see S1 and S2 for cluster results). Political journalists were coded into German media, foreign media, politics, NGOs, and political journalism. For data journalists, these categories were used: media, NGOs, data visualization advocates, politics, foreign media, data journalists, non-date journalists, and others.

In the second step, we extracted the hashtags used in the tweets across the data. These resulted in 32,200 hashtags for politics and 4,918 for data journalists. These were then clustered into categories (see S3 and S4 for cluster results). We used COVID-19, politics, elections, and climate for political journalists — a category for others was not required for the subset of hashtags. For data journalists, these were COVID-19, politics, elections, ddj (data-driven journalism), climate, sports, and others.

\section{Results}

Men are not just over-represented in our sample, they also tweet significantly more (724.47 Tweets per man / 289.23 Tweets per woman across both groups on average). Consequently, men created a large majority of tweets. Women wrote less than 19 percent of data journalist tweets and only 13 percent of political journalists' tweets (Table \ref{tab:descriptive-stats}). This is also consistent for data journalists, although not in a similar dimension. Men political journalists also use more mentions on average, measured by extracting all strings prefixed by an at-sign, which is Twitter's specification for tagging usernames. Women political and data journalists receive slightly more retweets on average.

\begin{table}[H] 
\caption{Summary Statistics of Gender, Retweets, and Mentions of Political (P) and Data (D) Journalists' Tweets\label{tab:descriptive-stats}}
\begin{tabularx}{\textwidth}{CCCCCCC}
\toprule
& & $n$ & $Share$ & $\overline{Tweets}$ & $\overline{Retweets}$ & $\overline{Mentions}$\\
\midrule
\multirow{2}{*}{P} & m & 375,582 & 0.87 & 4,803.0 & 248.6 & 1.14 \\
& f & 55,211 & 0.13 & 1,415.5 & 293.6 & 1.25 \\
\multirow{2}{*}{D} & m & 38,815 & 0.81 & 1,425.2 & 484.1 & 1.34 \\
& f & 8,997 & 0.19 & 1,358.0 & 506.0 & 1.41 \\
\bottomrule
\end{tabularx}
\end{table}

\subsection{Data journalists have a more open discourse.}

Part of our research focused on a general question about the arena of debate that takes part on Twitter. By extracting all mentions and comparing these users to our pre-compiled lists by cross-tabulation, we can show the share of references that stay within the political and data journalistic network.

Of all mentions by the political journalists, 10.7 percent are referenced within our sample, and 89.3 are outside of our sample. This number is even lower for data journalists. Only 8.2 percent of mentions are within the data journalistic community, and nearly 92 percent elsewhere. We find a statistically significant difference ($\chi^{2}$ = 429.06, p \textless  0.001, df = 1) between the two groups, with data journalists including more outsiders in their discourses, therefore being a less closed network compared to political journalists, which confirms H1.

\subsection{Women favor their peers in mentions.}

We have already shown in Table~\ref{tab:descriptive-stats} that there is a gap between the gender share of tweets and the gender share of users. This divergence can also be observed in the cross-tabulated share of mentions. This analysis only applies to tweets among our observed journalist users because we cannot derive gender for others.

Mentioning women users tend to favor their peers regarding mentions when tweeting in the journalistic bubble. Political journalists mention their peers in 27.4 percent of mentions, which is close to their share in the sample but more than their share on all tweets in the sample. This effect is even more extensive for mentioning women data journalists: they mentioned other women data journalists in 35.9 percent of intra-data journalistic discourses, which is even higher than the share of women in the sample.

\begin{quote}
    ``RT @mjKolly: Open question: How could and should people in the media industry credit each other's work?'' - @datentaeterin (2021-02-01 03:41:12pm)
\end{quote}
\begin{quote}
    ``RT @datentaeterin: "Anyone who wants to work in journalism should be able to handle data," says @ChElm in an interview with @journocode. That's why she wants to anchor data skills more firmly in education, for example, at the @IJ\_Online \#ddj'' - @daten\_drang (2021-10-06 07:37:58pm)
\end{quote}

Mentioning men, in comparison, only mentioned women in 17.0 percent for political journalists and 20.7 percent for data journalists; see Table~\ref{tab:mentioned}.
A Chi-Squared analysis showed a statistically significant result for both groups (p \textless 0.001). The effect size $\phi$ is 0.10 for political ($\chi^{2}$ = 549.78, df = 1) and 0.13 for data journalists ($\chi^{2}$ = 112.03, df = 1), which points to a small effect. A contribution analysis shows that the mention of women by women composes 66.48 percent of the measured effect for political journalists and 63.76 percent for data journalists. We are, therefore, able to confirm H2.

\begin{table}[H] 
\caption{Gender of mentioned users by author's gender\label{tab:mentioned}}
\begin{tabularx}{\textwidth}{CCCCCCC}
\toprule
& & \multicolumn{2}{c}{Mentioned} & \\
    \cmidrule{3-4}
    &        & m      & f &                                                   \\
    \midrule
    \multirow{2}{*}{Mentioning Political} & m   & 83.0\%    & 17.0\% & $\chi^{2}$ = 549.78, p \textless 0.01 \\
    & f & 72.6\%    & 27.4\% & \\
    \midrule
    \multirow{2}{*}{Mentioning Data} & m   & 79.3\%    & 20.7\% & $\chi^{2}$ = 112.03, p \textless 0.01 \\
    & f & 64.1	\%    & 35.9\% & \\
\bottomrule
\end{tabularx}
\end{table}

\subsection{Retweets are more evenly distributed for data journalists.}
\begin{figure}[H]
\centering
\includegraphics[width=\linewidth]{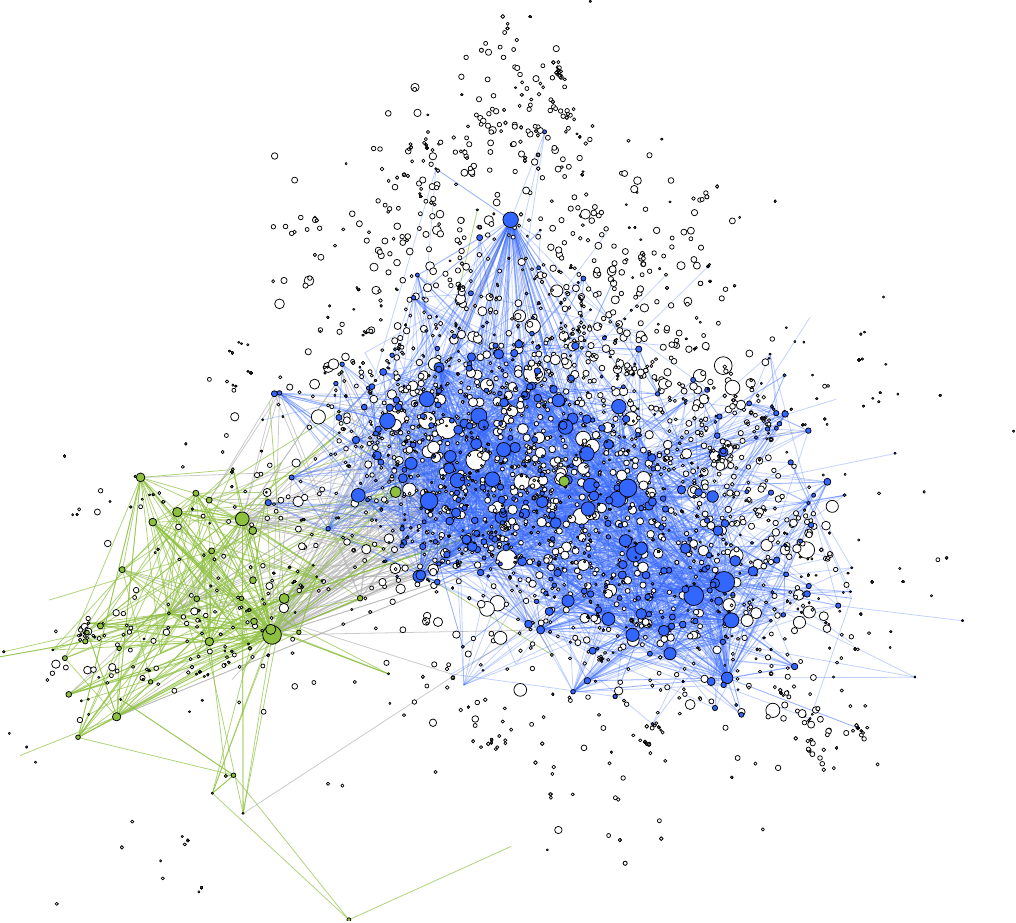}
\caption{Retweet-network of political and data journalists: showing edges when at least 2 retweets were sent, node sizes by degrees. Green nodes and edges: data journalists, blue nodes and edges: political journalists, white nodes: unknown (users were not part of pre-defined lists), gray edges: connections between political and data journalists.\label{fig:graph_retweets_ddj_pol}}
\end{figure}

While mentions are unevenly shared between genders in both groups, this is not identical concerning retweets. Women political journalists are only retweeted by men in 13.3 percent of intra-journalistic retweets; men data journalists only share tweets of women in 18.4 percent of cases, as shown in Table \ref{tab:retweeted}. Again, the share of women retweeting other women is higher in both groups but lower than their share of users in both cases. Pearson's Chi-squared test shows a statistically significant result for political journalists. For data journalists, the results are not significant. The effect size of 0.10 is small for political journalists and even smaller for data journalists. While residues and contributions favor an effect between women for political journalists, this is not the case for data journalists. The effect on them seems to be much smaller. H3 can be confirmed very certainly for political journalists but not for data journalists. See Figure \ref{fig:graph_retweets_ddj_pol} for a full-size network of political and data journalists' retweets.

\begin{table}[H] 
\caption{Gender of retweeted user by author's gender\label{tab:retweeted}}
\begin{tabularx}{\textwidth}{CCCCC}
\toprule
& & \multicolumn{2}{c}{Retweeted} & \\
    \cmidrule{3-4}
    &        & m      & f &                                                   \\
    \midrule    
    \multirow{2}{*}{Retweeting Political} & m   & 86.7\%    & 13.3\% & $\chi^{2}$ = 326.42, p \textless 0.01 \\
    & f & 76.5\%    & 23.5\% & \\
    \midrule    
    \multirow{2}{*}{Retweeting Data} & m   & 81.7	\%    & 18.3\% & $\chi^{2}$ = 1.56, p \textgreater 0.05 \\
    & f & 78.5\%    & 21.5\% & \\
\bottomrule
\end{tabularx}
\end{table}

\unskip

\subsection{Sources are different between sexes — and journalistic disciplines.}

To analyze the references on sources, we compared all usernames retweeted by our population. 

When analyzing retweeted accounts, we find that only a part of the sources is identical across sexes for both groups of journalists. Political journalists have used 20,937 users as sources; 19 percent are common across both sexes for political journalists and around 14 percent for data journalists — of a total of 6,519, confirming H4. 

We also find differences in self-sourcing shares, where users retweet their own tweets for their audience.
The share of self-retweets is 2.57 times significantly higher for political journalists than for data journalists. In particular, male political journalists significantly self-retweet themselves a lot more than their female colleagues, whereas this finding is exactly the opposite for data journalists — although in a much smaller size. 
A logistic regression was used to analyze the relationship between sex, the area of journalism, and a self-retweet. It was found that holding all other predictor variables constant, the odds ratio of a self-retweet occurring increased on average by 6.56 (95\% CI 5.19, 8.3) for the occurrence of male sex. It was also found that, under the same conditions, the odds ratio of a self-retweet occurring increased on average by 2.43 (95\% CI 2.02, 2.93) for an occurrence of political journalism.

To further understand possible clusters of sources, we added a content analysis at this stage:

First, we manually coded the top 30 retweeted sources for each sex and field.
While political journalists used German media and other political journalists as their primary sources across the sexes, with a few men referring to foreign media, data journalists also leverage non-peer journalists in their retweets. Women data journalists also used non-governmental actors or data visualization advocates (private companies) as sources.

Second, an analysis of the 30 most used hashtags revealed little differences for political journalists but a much more diverse set of topics for data journalists, confirming H5 — with women seemingly sharing different topics than their men counterparts (see Tables \ref{tab:top30_ht_pol} and \ref{tab:top30_ht_ddj}).

\begin{table}[H] 
\caption{Shares of Clusters of Top 30 hashtags by sex of Political Journalists.\label{tab:top30_ht_pol}}
\begin{tabularx}{\textwidth}{CCC}
\toprule
        \textbf{Cluster} & \textbf{f} & \textbf{m}\\
        \midrule
        Politics & 51.94 & 48.88\\
        Covid-19 & 28.77 & 37.04\\
        Elections & 15.38 & 14.08\\
        Climate & 3.92 & -\\
\bottomrule
\end{tabularx}
\end{table}

\begin{table}[H] 
\caption{Shares of Clusters of Top 30 hashtags by sex of Data Journalists.\label{tab:top30_ht_ddj}}
\begin{tabularx}{\textwidth}{CCC}
\toprule
        \textbf{Cluster} & \textbf{f} & \textbf{m}\\
        \midrule
        Data journalism & 39.46 & 33.05\\
        Covid-19 & 17.77 & 36.67\\
        Politics & 15.21 & 13.22\\
        Elections & 15.06 & 9.59\\
        Climate & 1.96 & 1.99\\
        Sports & - & 1.24\\
        Others & 10.54 & 4.23\\
\bottomrule
\end{tabularx}
\end{table}

\section{Networks}

To be able to confirm our insights and show the utility of network analysis for this task, we modeled the data as four distinct networks for each profession: a profession-internal retweet network, a retweet network that incorporates all internal and external retweets, an internal network of mentions, and a network of hashtags used in tweets.

\subsection{Internal Retweets and Mentions} To further understand the dynamics of retweets, we created a retweet network. Purple nodes represent women, and green nodes represent men. Gray edges represent at least two retweets between men in both directions, purple edges at least two retweets between women, and green edges a retweet connection between a man and a woman user. For data journalists, we show edges for at least one retweet in both directions, as the network is much smaller. 

\begin{figure}[H]
\centering
\includegraphics[width=10.5cm]{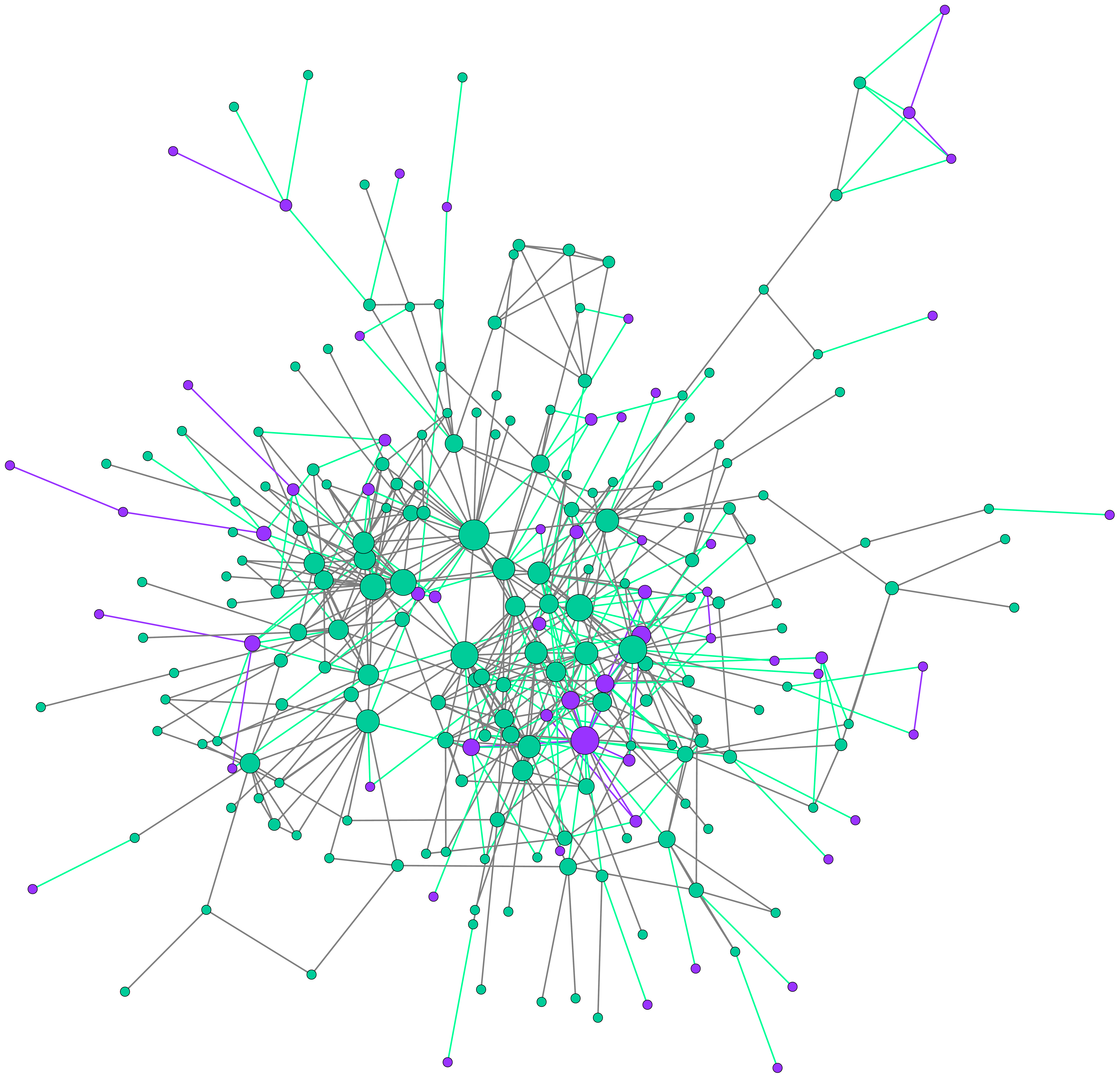}
\caption{Retweet-network of political journalists by gender: Graph network of retweets by German political journalists, showing edges when both nodes sent at least 2 mutual retweets, node sizes by degrees. Purple nodes: women, green nodes: men, green edges: men-women, purple edges: both women, gray edges: both men.\label{fig:graph_retweets_pol}}
\end{figure}

While women networks are hard to spot in the network of political journalists (see Fig.~\ref{fig:graph_retweets_pol}), we find clusters of affiliations between different publishers. While reporters and editors for the media company \textit{Axel Springer} and its outlets are closely connected on the left, journalists for \textit{Der Spiegel} or \textit{S{\"u}ddeutsche Zeitung} are found on the lower right.

\begin{figure}[H]
\centering
\includegraphics[width=10.5cm]{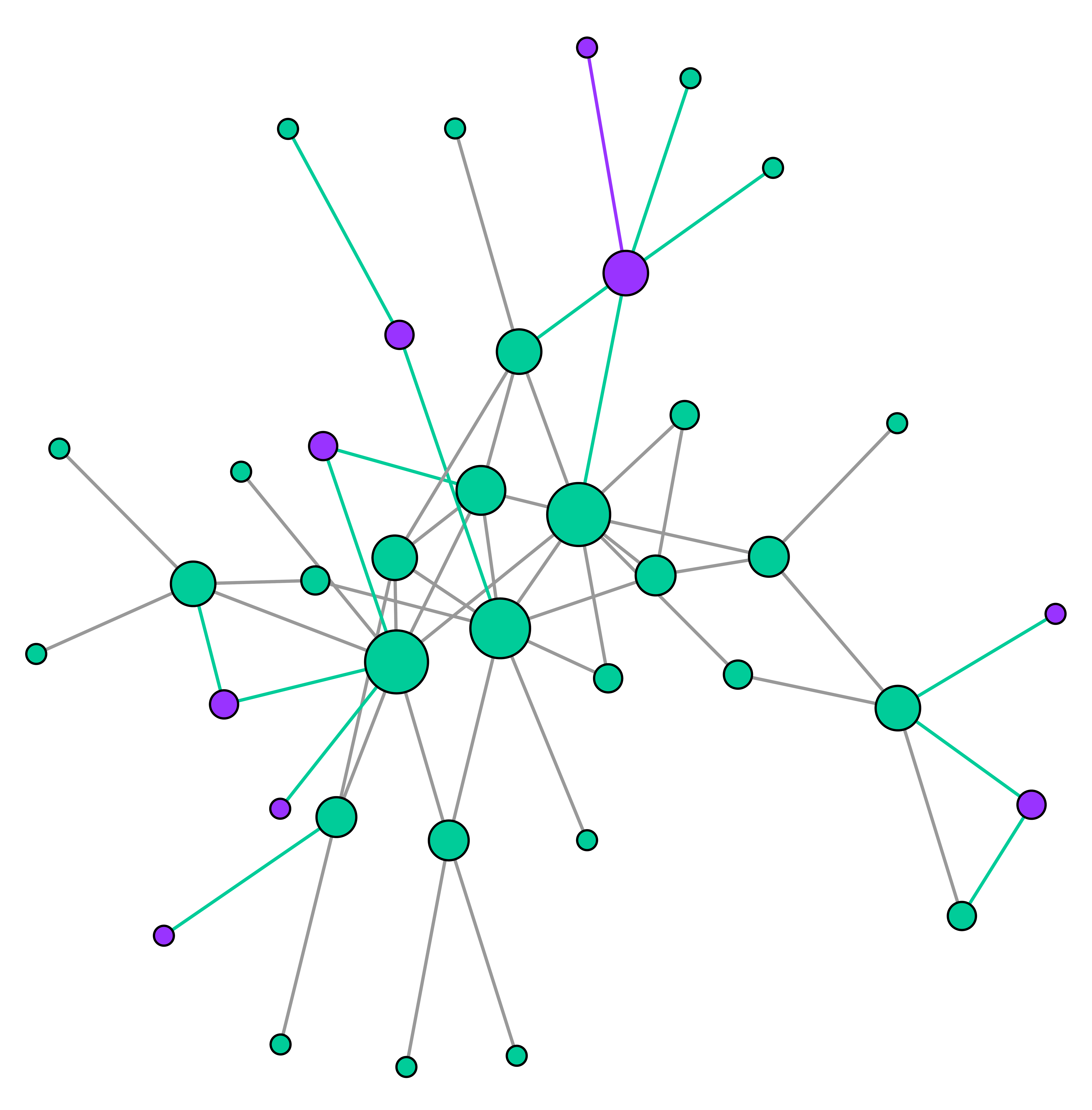}
\caption{Retweet-network of data journalists by gender: Graph network of retweets by German data journalists, showing edges when both nodes sent at least one mutual retweet, node sizes by degrees. Purple nodes: women, green nodes: men, green edges: men-women, purple edges: both women, gray edges: both men.\label{fig:graph_retweets_ddj}}
\end{figure}

The data journalists' network does not show similar patterns, which the smaller team sizes in the field could influence (see Figure ~\ref{fig:graph_retweets_ddj}).

We found the values of in-degree ($t(642) = 2.0341, p < 0.05$), out-degree ($t(440.09) = 2.9155, p < 0.01$), and Kleinberg's authority centrality score ($t(642) = 2.0108, p < 0.05$) to be statistically significant for political journalists, while they are not for data journalists. See Tables \ref{tab:mentionsProperties} and \ref{tab:retweetsProperties} for network property metrics.

\begin{table}[H] 
\caption{Properties for networks of mentions between political (P) and data journalists (D).\label{tab:mentionsProperties}}
\begin{tabularx}{\textwidth}{CCC}
\toprule
        \textbf{Property} & \textbf{P} & \textbf{D} \\
        \midrule
        Diameter & 8.00 & 6.00\\
        Mean Dist. & 2.81 & 2.34\\
        Edge Density & 0.03 & 0.088\\
        Reciprocity & 0.48 & 0.53\\
        Transitivity & 0.28 & 0.41\\
        Components & 1.00 & 6.00\\
\bottomrule
\end{tabularx}
\end{table}

\begin{table}[H] 
\caption{Properties for networks of retweets between political (P) and data journalists (D).\label{tab:retweetsProperties}}
\begin{tabularx}{\textwidth}{CCC}
\toprule
        \textbf{Property} & \textbf{P} & \textbf{D} \\
        \midrule
        Diameter & 8.00 & 7.00\\
        Mean Dist. & 3.05 & 2.76\\
        Edge Density & 0.02 & 0.05\\
        Reciprocity & 0.33 & 0.28\\
        Transitivity & 0.26 & 0.33\\
        Components & 1.00 & 1.00\\
\bottomrule
\end{tabularx}
\end{table}

We further analyzed community structures by running community detection algorithms. Louvain clustering \citep{Blondel2008} reached the highest modularity of 0.53. The algorithms seek to optimize by increasing the relative density of edges inside communities compared to outside communities. While we find more communities for political journalists --- plausible, due to the higher number of tweets and users, we find that the median share and counts of women in the communities are higher for data journalists in mention networks (see Table \ref{tab:communityStats}).

\begin{table}[H] 
\caption{Summary Statistics for Louvain community detection.\label{tab:communityStats}}
\begin{tabularx}{\textwidth}{CCCCCC}
\toprule
Graph & Count & $\overline{Users}$ & $\overline{Share_W}$ & $med$ & $\overline{Women}$ \\
\midrule
Mentions D & 37 & 3.62 & 0.43 & 0.25 & 1.11\\
Mentions P & 135 & 4.77 & 0.36 & 0.00 & 1.33\\
RTs D & 69 & 1.68 & 0.30 & 0.00 & 0.49\\
RTs P & 202 & 2.82 & 0.28 & 0.00 & 0.75\\
\bottomrule
\end{tabularx}
\end{table}

\subsection{External Retweets}

Creating networks of sources for both sexes and both professions leads to further conclusions. We compared both sexes and areas of journalism in all their retweeted messages. This also included outsiders of their journalistic circles, different from the analysis above. The intent was not only to understand the journalistic communities' internal structures but also their differences in leveraging different external actors. The network was constructed by defining all nodes as users and tweets retweeting each user as edges.

Women source networks for political journalists have a larger diameter and a higher mean distance between nodes than men political journalists (see Table \ref{tab:sourcesProperties}). That indicates that these networks are further spread out, while women data journalists form a much more compact sources network.

Reciprocity numbers indicate a more coherent sourcing behavior for political journalists, who seem to reference themselves more than data journalists (as described above). This can also be seen in transitivity metrics describing how likely adjacent nodes are connected, revealing tightly connected communities. Political journalists of both sexes have metrics an order of magnitude higher than data journalists.

\begin{table}[H] 
\caption{Properties for networks of sources for Political and Data journalists of women (F) and men (M).\label{tab:sourcesProperties}}
\begin{tabularx}{\textwidth}{CCCCC}
\toprule
& \multicolumn{2}{c}{Political} & \multicolumn{2}{c}{Data} \\
            \cmidrule(lr){2-3} \cmidrule(lr){4-5}
            Property     & F             & M             & F           & M          \\
            \midrule
            Diameter     & 13.00          & 9.00           & 6.00         & 9.00        \\
            Mean Dist.   & 4.69           & 3.57           & 2.50         & 3.46        \\
            Reciprocity  & 0.016         & 0.036         & 0.0026      & 0.0217     \\
            Transitivity & 0.009         & 0.017         & 0.0012      & 0.0079     \\
            Components   & 4.00           & 2.00           & 1.00         & 1.00        \\
            Mean Indegrees  & 556.31        & 2863.24       & 492.09      & 632.40     \\
            Mean Outdegrees & 149.73        & 239.31        & 19.54       & 24.03      \\
\bottomrule
\end{tabularx}
\end{table}

\subsection{Hashtags}

Lastly, a two-mode network of hashtags helps to deepen the understanding of the diversity of topics that the journalistic domains mention in their tweets. This network was created by defining hashtags and users as nodes and tweets containing a certain hashtag as edges.

While we have found a consensus within the most prevalent hashtags for political journalists above, we can see that women political journalists form a hashtag network with nine components, which indicates that there are separate, unconnected parts of the network, which we can identify as users that do not use hashtags at all. This is also observable for men political journalists, but only with two components, not for data journalists, who seem to form a joint network connected via shared hashtags.

We use the average ratio $\frac{Used Hashtags By User}{Hashtag Used By Users}$ to compare groups for an indication of the prevalence of hashtag use. It's important to note that in- and outdegrees are not typically reported for two-mode networks. However, we have included this comparison in this particular case to provide a more comprehensive analysis. We find that women political journalists, on average, create 285.12 outdegrees, with hashtags averaging 34.75 indegrees, on average, a ratio of 108.46. On the contrary, men political journalists have an average of 679.95 outdegrees, their hashtags' degrees being on a similar level to women with 27.52 indegrees, a ratio of 280.01. Data journalists have much lower values, with women creating on average 138.31 outdegrees and 6.46 indegrees (ratio: 88.46), and males ending up with 178.23 out- and 6.43 indegrees, and a ratio of 108.35. This shows that political journalists use, on average, much more hashtags than data journalists, with men being ahead in each case.

\section{Discussion}

We have shown differences in mention and retweeting behavior between the sexes of political and data journalists in Germany, confirming H2 and H3 for political journalists. 

Women tend to mention their peers more often in tweets than men. Since men comprise the larger share of the Twitter network, they tend to be more visible. The differences could make women and their work less apparent on Twitter, therefore receiving less amplification and legitimization. This work contributes a non-US perspective on differences in Twitter communication styles between sexes and different groups of journalists. The results indicate existing norms within newly created public communication spheres, a selective, gatekeeping process on the disseminator side of information \citep{White1950}.

This effect can also be found in retweets of political journalists, although it is not similarly strong for data journalists. This might indicate that data journalists tend to share the work of others with less regard for the sexes compared to their colleagues in political reporting. However, women journalists show higher rates of retweet behavior than their peers. Since the effect appears in both groups, this indicates that women pay greater attention to tweets by other women; however, this effect is much more solid for political journalists. These results confirm earlier work by \citep{Usher2018}. 

Although their share of identical sources across sexes is low, political journalists tend to focus their retweets of central issues mainly on direct peers or media sources, while data journalists seem to convey their most prevalent sources from a more diverse spectrum of backgrounds and less from other data journalists. This might indicate the broader background that data journalism has as a discipline and again points to the homophilous network structures of political journalists that have been described before \citep{Hanusch2019,Molyneux2015,Molyneux2019,Mourao2015,Mourao2016,Nuernbergk2016}. However, this finding needs to be regarded in combination with a general contrast between sexes, which points to the fact that women and men journalists retweet different voices on Twitter. 

Even women political journalists' sources networks are larger and might incorporate a more diverse set of sources than men's. There seems to be a contrasting finding regarding the higher number of average hashtags for political journalists, which might indicate a higher diversity of topics for this area but could also point to the fact that hashtags emerge quickly for breaking-news political events rather than for data journalists topics that are rarely reporting news up-to-the-minute \citep{Vicari2018,Lin2021,Zhang2017a}, as we do not find indications for this when analyzing the most prevalent hashtags. This, however, might be concealed for this method and is a starting point for further research.

While analyzing retweet networks, we found visual evidence of clusters of affiliations that might impact tweet behavior, which might also be a vantage point for further research. Community detection found that data journalists' mention networks tend to have a higher median share and count for women than other studied networks.

In contrast to earlier findings, we found relatively high percentages for external mentions and retweets for both journalistic groups. \citet{Nuernbergk2016} had shown a journalistic-political Twittersphere, which mainly referenced each other, similar findings by \citet{Mourao2015} and \citet{Molyneux2019}.
However, we need to point out that as this work did not include a broad set of politicians in the sample or control for different groups of non-journalistic actors apart from highlighting the most common retweeted users and focused more on gender differences, the results are not fully comparable with those earlier results. We also did not include media companies' Twitter accounts in our analysis, which might make up a large share of mentioned or retweeted users, as seen by visual inspection.

We also need to clarify that the networks we were investigating were those formed by retweets and mentions to create inner social Media amplification of messages, ignoring the latent network of followers and following relations that might lead to messages being transported outside the Social Media network. Our research is further limited by the influence of Twitter's algorithms, which might have shaped the tweets shown to users. However, Twitter was still defaulting to a chronological order in displaying tweets at the time of analysis. The study focuses exclusively on German journalists, which restricts the generalizability of the findings to other countries or cultures. 

As often with quantitative methods, they lack depth in understanding the reasons behind the observed behaviors. Qualitative methods, such as structured interviews, could provide further research opportunities into the motivations, perceptions, and challenges journalists face on Twitter. This could enhance the understanding of how gender dynamics manifest in the digital interactions of journalists.

We could, however, find a significant difference in shares of those internal discourses between the two groups of journalists, with data journalists being less locked than political journalists. That might indicate a greater openness to the influence of others in the data journalistic community, which is a finding that could need closer examination.

\section{Conclusions}

We compared the Twitter networks of German political and data journalists to analyze differences in communication between women and men. We did find a difference in the proportions of internal discourses within the two groups of journalists. Data journalists tended to have fewer internal discussions on Twitter than political journalists. However, we could not reproduce earlier findings, which showed an elitist network of political journalists on Twitter.

This study showed that men dominated the number and share of tweets in networks of political and data journalists in Germany. While women were much less mentioned and retweeted by men, other women tended to favor their peers. This effect was visible in both groups for mentions and was also observable for retweets by political journalists and, to a lesser degree, for data journalists. This indicates a different perception of the work and arguments made by colleagues on Twitter between genders, which might lead to less amplification and legitimization of women's voices on Twitter. Further research is required to extract the causes behind this effect and the possibilities of countering this behavior.

\vspace{6pt} 

\section*{Acknowledgements}

\subsection*{Contributions}
Conceptualization, B.W. and J.P.; Data curation, B.W.; Formal analysis, B.W.; Investigation, B.W.; Methodology, B.W.; Project administration, B.W.; Software, B.W.; Supervision, J.P.; Validation, J.P.; Visualization, J.P.; Writing – original draft, B.W. and J.P.

\subsection*{Funding}
This research received no external funding

\subsection*{Data Availability}

Data are available upon request.

\subsection*{Acknowledgments}

The authors would like to thank Rebecca Leipsic for helping to compile the list of political journalists.

\subsection*{Conflicts of Interest}

The authors declare that they have no conflict of interest.

\subsection*{Supplementary Material}

The following supplementary material is available: Table S1: Clusters of Sources: Political Journalists, Table S2: Clusters of Sources: Data Journalists, Table S3: Cluster of Hashtags: Political Journalists, Table S4: Cluster of Hashtags: Data Journalists

\end{document}